\documentclass[preprint,showpacs,preprintnumbers,amsmath,amssymb]{revtex4}
\usepackage{amsmath,amssymb,graphicx} 
\usepackage{epstopdf}
\usepackage {amssymb}
\newcommand{\nc}{\newcommand}
\nc{\ba}{\begin{eqnarray}}
\nc{\ea}{\end{eqnarray}}
\newcommand\bb{{\mathbf{b}}}
\newcommand\A{{\mathbf{a}}}
\newcommand\s{\sigma}

\nc{\ga}{\gamma}
\nc{\x}{{\bf x }}
\nc{\kk}{{\bf k }}
\nc{\ds}{{\dot {s}_3}}

\input{epsf.sty}

\begin{document}

\preprint{CERN-PH-TH/2009-140}

\title{ Zipping and Unzipping of Cosmic String Loops in Collision }

\author{ H. Firouzjahi} \email{firouz@ipm.ir}
\affiliation{ School of Physics, Institute for Research in Fundamental Sciences (IPM), \\ P .O. Box 19395-5531,Tehran, Iran
}

\author{J. Karouby }\email{karoubyj@physics.mcgill.ca}
\affiliation{\,  Physics Department, McGill University,   Montreal, H3A 2T8, 
Canada. }

\author{S. Khosravi} \email{khosravi@ipm.ir}
\affiliation{ Physics Department, Faculty of Science, Tarbiat Mo'alem University, Tehran, Iran, and  School of Astronomy, Institute for Research in Fundamental Sciences(IPM), Tehran, Iran.
}

\author{R. Brandenberger}  \email{rhb@physics.mcgill.ca}
\affiliation{ Physics Department, McGill University,   Montreal, H3A 2T8, 
Canada., and
Theory Division, CERN, CH-1211 Gen\`eve, Switzerland}


\begin{abstract}
In this paper the collision of two cosmic string loops is studied. 
After collision junctions are formed and the loops are entangled. We show that
after their formation the junctions start to unzip and the loops disentangle.
This analysis provides a theoretical understanding of the unzipping effect  observed in 
numerical simulations of a network of cosmic strings with more than one type of cosmic strings.
The unzipping phenomena have important effects in the evolution of cosmic string networks  
when junctions are formed upon collision, such as in a network of cosmic superstrings.


\end{abstract}

\maketitle

\section{Introduction}

Cosmic strings are copiously produced at the end of brane inflation 
\cite{Sarangi:2002yt, Majumdar:2002hy} (for reviews see e.g. 
\cite{HenryTye:2006uv, Kibble:2004hq, ACD, Mairi}). These strings are in the form of 
fundamental strings (F-strings), D1-branes (D-strings) or their bound states. F- and D-strings 
can combine to form bound states - $(p,q)$ strings - which are constructed from 
$p \,$ F-strings and $q \, $D-strings on top of each other. Due to charge conservation, when  
two $(p,q)$ cosmic strings intersect generally a junction is formed. This is in contrast to 
what happens in the case of U(1) gauge cosmic strings: When two U(1) gauge cosmic strings 
intersect, they usually exchange partners and intercommute with probability close to unity. 
In this view, the formation of junctions may be considered as a  novel feature of the network of 
cosmic superstrings.  Networks of strings with junctions have interesting physical properties, 
such as the formation of multiple images \cite{Shlaer:2005ry, Brandenberger:2007ae}
and non-trivial gravity wave emission \cite{Brandenberger:2008ni, Leblond:2009fq}.
Different theoretical aspects of $(p,q)$ string construction were studied in  \cite{Copeland:2003bj, Firouzjahi:2006vp, Jackson:2004zg, Firouzjahi:2007dp, Cui:2007js, Davis:2008kg}
while the cosmological evolution of a string network  with junctions has been investigated in 
\cite{Tye:2005fn}. 

The evolution of a network containing two types of cosmic strings
was studied by Urrestilla and Vilenkin \cite{Urrestilla:2007yw}. In their model, the cosmic strings 
are two types of U(1) gauge strings with interactions between them. Let us label these strings 
as A and B strings. Due to the interaction, the strings cannot exchange partners and a bound state, 
string AB,  will form if the strings are not moving too fast.
It was shown that the length and the distribution of the string network are dominated by the original 
A and B strings and there is a negligible contribution to the string network length and population 
from the bound states strings AB. This can be understood based on the following two reasons. Firstly, 
the junctions may not form if the colliding strings are moving very fast  so  they can simply pass through 
each other \cite{Bettencourt:1994kc, Copeland:2006eh, Copeland:2006if, Copeland:2007nv, Salmi:2007ah, Bevis:2008hg}. Secondly and more 
curiously, if the junctions are formed, they start to unzip during the 
evolution. The process of zipping and unzipping of cosmic strings in collision is a
non-trivial dynamical property. Our aim here is to provide some theoretical understanding
of how this process happens in the collision of cosmic strings loops.

\section{The Setup}

Here we provide our setup. We consider two cosmic string loops moving in
opposite directions. At the time of the collision, junctions are formed. This can be viewed as a
generalization of straight strings collision 
\cite{Copeland:2006eh, Copeland:2006if, Copeland:2007nv}.
However, due to topological constraints,  there are new non-trivial effects which 
can lead to the unzipping of junctions. This is unlike what happens in the case of straight strings,
where for two colliding straight strings, once the junctions are formed, they will always grow with time and do not unzip \cite{Copeland:2006eh, Copeland:2006if, Copeland:2007nv}.

\begin{figure}[t] 
\vspace{-2.8cm}
   \centering
    \includegraphics[width=6in]{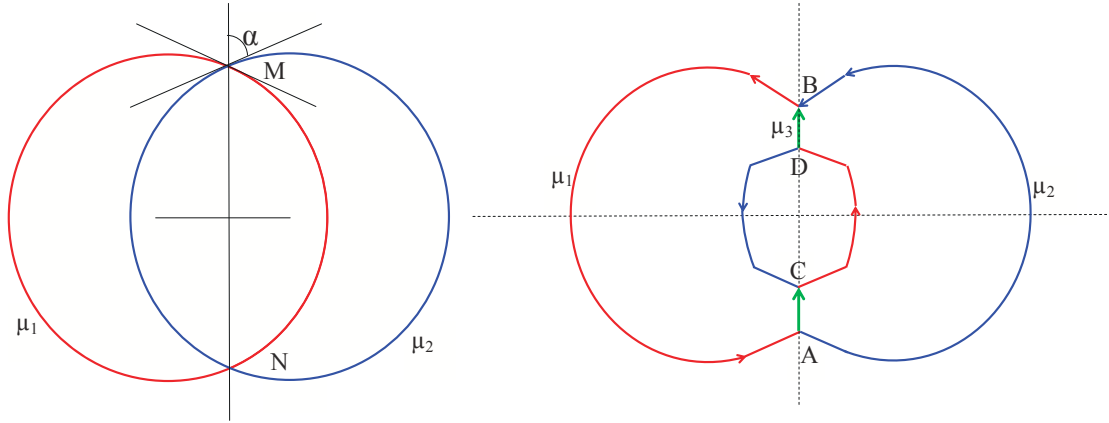}
    \hspace{0cm}
    \vspace{-3cm}
\caption{ A schematic view of the loops at the time of collision(left) and after collision (right). The 
arrows in the right figure indicate the directions in which the $\sigma_i$ coordinate increases. 
We use the convention that on a loop $\s_i$ runs counter clockwise.  There are four junctions 
and eight kinks in total. 
 }
\label{collision}
\end{figure}

In order to simplify the analysis, we assume the colliding loops have equal tensions and radii,
that the planes they span are parallel. Choosing the center of mass frame, we take the loops 
to be in the $x-y$ plane and assume they are moving along the $z$-axis with speed $\pm v$. 
A schematic view of the collision is shown in {\bf Fig. \ref{collision}}. The collision happens at 
$t=0, z=0$. There are two collision points. The angle of collision, $2 \alpha$, is defined as the 
angle between the tangential lines to the loops at the points of collision.

We choose the incoming strings to be of the form of simple loops in their rest frames
\ba
\label{Radius}
R(t) = R_0 \cos \left(\frac{t-t_0}{R_0} \right) \, .
\ea
This may not be a realistic configuration, but due to the complexity of the collision analysis, this
ansatz is illustrative enough to capture the unzipping effect. Here $t_0$ is the phase at the time
of collision and the radius at collision time is $R_0 \,  \cos(t_0/R_0)$. 

After the collision, four junctions are formed (in {\bf Fig. \ref{collision}} they are denoted by  
$A, B, C$ and $D$). Due to symmetry, one expects that junctions A and B and junctions C and D 
evolve similarly but in opposite directions.. On each junction, there are three string legs,
 two of them are the incoming strings and the third is the newly formed string with tension
 $\mu_3$. As explained above, we assume that the incoming loops have equal tensions, 
 $\mu_1=\mu_2$. Later we will see that for a junction to form one requires that 
 $2 \mu_1 > \gamma \mu_3$.

Due to the symmetry, the third string  is stationary and 
is oriented either along the y-axis (y-link) or x-axis (x-link). The
orientation of the third string is controlled by the angle $\alpha$. For small $\alpha$ (roughly
$0<\alpha< \pi/4$) we expect a y-link junction and for a larger value of $\alpha$ an x-link 
junction. For the discussion below and in {\bf Fig. \ref{collision}} we consider
a y-link junction.

Guided by the causality and the symmetry of the problem, one expects that, after collision, 
the entangled loops are divided into two secondary loops, the external loop and the internal 
loop. The external and the internal loops are connected by the newly formed strings with
tension $\mu_3$. Given the symmetry of the setup, a nice feature of the internal and external 
loops are that half of each is from the first string and the other half is from the second string. There 
are four kinks on each secondary loop separating the newly formed arcs from the parts of the old 
loops which do not yet feel the presence of the junctions (by causality).
As we shall see, these non-trivial topological constraints between the internal and external loops
play an essential role in the unzipping process.

At the time of collision the system has a non-zero angular momentum around the axis of collision.
However, we do not expect the angular momentum to play an important role in the
unzipping process. As we shall see below, the unzipping process is determined by forces in
the place of the strings, whereas angular momentum induces forces in orthogonal direction.
In addition, these forces will vanish at the local of the junctions. 

The world-sheet of each string is described by a temporal coordinate $ \tau $ and
a string length parameter $\s_i$.  We take each string to have its own $\sigma_i$ parameter.
Our convention  for the orientation of $\sigma_i$  is that on a given loop, whether an original 
colliding loop or a secondary loop, the $\sigma_i$ coordinate increases counter-clockwise
from $0$ to $2 \pi$. For example, at the time of collision, the point $M$  in  {\bf Fig. \ref{collision}} 
has $\sigma_1=  \gamma R_0 \alpha  $ on string 1 and $ \sigma_2=  \gamma R_0 (\pi - \alpha) $
on string 2. Similarly, the point N has $\sigma_1= \gamma R_0 (2 \pi -\alpha)  $ on string 1 and 
$\sigma_2 = \gamma R_0(\pi + \alpha) $ on string 2. Here $\gamma$ is the Lorentz factor,
$\gamma^{-2} = 1- v^2$, which shows up due to the boost from the string rest frame to the
center of mass frame. Finally on the third string, $\sigma_3$ increases from south to north.
 
One complexity of dealing with loops in collision is the orientation of the $\sigma_i$ coordinate
at junctions. We follow the prescription of \cite{Bevis:2009az} and use the sign parameterization 
for $\delta_i$ according to which $\delta_i$ can take values $\pm 1$. If the value of $\sigma_i$
of a particular string increases(decreases) towards the junction, we assign $\delta_i=+1(\delta_i=-1)$.
With this prescription, the two ends of a piece of string ending in two neighboring junctions have
opposite $\delta$ parameters. The arrows in  {\bf Fig. \ref{collision}} indicate this prescription. 
Since it is important for the later analysis, we now give the values of $\delta_i$ at each junction:
\ba
A:  \left| 
\begin{array}{c}
\delta_1=+1 \\
\delta_2 =-1\\
\delta_3=-1
\end{array} 
\right. \hspace{0.5cm}
B:  \left| 
\begin{array}{c}
\delta_1=-1 \\
\delta_2 =+1\\
\delta_3=+1
\end{array} 
\right.
 \hspace{0.5cm}
C:  \left| 
\begin{array}{c}
\delta_1=-1 \\
\delta_2 =+1\\
\delta_3=+1
\end{array} 
\right. \hspace{0.5cm}
D:  \left| 
\begin{array}{c}
\delta_1=+1 \\
\delta_2 =-1\\
\delta_3=-1
\end{array} 
\right.
\ea

We consider a flat space-time background. The induced metric $\ga_{i\, ab}$ on each string is given 
by 
\ba
\label{gamma}
\ga_{i\, ab} = \eta_{\mu \nu}  \,  \partial_{a } \,  x_{i}^{\mu} \partial_{b} \, x_{i}^{\nu}  \, .
\ea
Here and in the following, we reserve $\{a, b\}= \{ \tau, \sigma_i \}$ for the string world-sheet indices 
while Greek indices represent the four-dimensional space-time coordinates. 
Furthermore, $x_{i}^{\mu}$ stands for the position of the $i$-th string in the target space-time.

After collision, the junction points correspond to the intersection of three segments of strings :
two from the colliding loops and one from the bound state string that appears after collision. 
Including the $\delta$-parameterization on each segment of strings, the Nambo-Goto action describing the dynamics of the strings positions, $x^\mu_i$, and the evolution of junction points  is \cite{Bevis:2009az}
\ba
\label{action}
S= - \sum_i \mu_i \delta_i \int d \tau \int d\s_i \sqrt{- {x'_i}^2 {\dot x_i}^2   } \, 
 \theta( s_i^{B_i}(\tau) - \sigma_i ) \,  \theta( -s_i^{A_i}(\tau) + \sigma_i ) \nonumber\\
 + \sum_i \sum_J \int d\tau f_{i \mu}^J . \left[  x_i^{\mu} (\tau, s_i^J (\tau)) - X^{\mu}_J(\tau) 
 \right] \, .
\ea
Here an over-dot and a prime denote derivatives with respect to $\tau$ and $\s$, respectively. 
The function $s_i^J(\tau)$ indicates the value of the $\s_i$ coordinate for the $i$-th
string at the junction $J$. 
The theta functions indicate the fact that each piece of string exists only between
junctions $B_i$ and $A_i$. In this notation, the $\sigma_i$ coordinate for the piece of the string 
which stretches from junctions $A_i$ to $B_i$ 
is increasing from $A_i$ to $B_i$ and 
\ba
s_i^{A_i}(\tau) \leq \sigma_i \leq s_i^{B_i}(\tau) \, .
\ea
 It should be noted that  in our case $\{A_i , B_i\}$ collectively stand for the junction points $J$
 with
 \ba
 J \in \{ A, B, C ,D\}
 \ea
 in {\bf Fig. \ref{collision}}.
Finally, the functions $f_{i\, \mu}^J$ are the Lagrange multipliers which enforce the constraints that at the junction $J$, the three strings meet and 
\ba
x_i(s_i^J(\tau) , \tau)= X^J(\tau) \, ,
\ea
where $X^J(\tau)$ is the junction position in target space-time.

As explained above, the value of the $\sigma_i$ coordinate for the $i$-th string at junction $J$
is given by the function $s_i^J(\tau)$. It is a dynamical variable which controls the evolution of
the junction. For example at junction $B$ in {\bf Fig. \ref{collision}} the process of zipping for  the string $\mu_3$ happens when $\dot s_3^B(\tau)>0 $ whereas its unzipping happens
when $\dot s_3^B(\tau)$ vanishes at some time  during evolution and $\dot s_3^B(\tau)<0$ afterwards.
Our goal in next section is to find the dynamical equations for ${\dot {s_i}}^{J}$ to understand
the process of zipping and unzipping of strings in junctions.

The derivation  of the equations of motion coming from action (\ref{action}) is given in 
\cite{Bevis:2009az}. Here we summarize the equations which are necessary for our colliding 
loop analysis.

We impose the conformal temporal gauge on the string world-sheet, namely 
$X_{i}^{0}= t=\tau$ and $\ga_{ i\, 0\s}=0$. This is equivalent to
\ba
\label{gauge1}
\dot \x_{i}  \, . \,  \x_{i}' =0 \quad , \quad \dot \x_{i}^{2} + \x_{i}'^{2} =1 \, 
\ea
where  the $\x_{i}$ represent the spatial components of the $i$-th string. The solution of string equation of motion, $\ddot \x_i - \x_i''=0$, as usual,  is given in terms of
left- and right-mover waves,
\ba
\label{right-left}
\x_{i}(t, \sigma) = \frac{1}{2}\,  \A_{i} (\frac{\s+t}{2}) +  \frac{1}{2} \, \bb_{i}(\frac{\s -t}{2})
\ea
 with ${\A'_i}^2={\bb'_i}^2=1 $. Imposing the junction conditions obtained from
varying the action (\ref{action}), one can 
find expressions for  $\A'_i$ and $\bb'_i$ at the position of the junction $J$. Imposing the conditions
${\A'_i}^2={\bb'_i}^2=1 $ one finds the following equations for $s_i^J$
\ba
\label{s-equation}
 \dot{s_i}^J = \delta_i \left( 1- \frac{  \mu M_i (1- c_i^J)  }{ \mu_i  \sum_k M_k (1- c_k)  } 
 \right)
\ea
where
$ \mu \equiv \sum_i \mu_i   \quad , \quad  $ and
\ba
M_i = \mu_1^2 - (\mu_j - \mu_k)^2    \quad \, \quad c_i^J(t) = Y_i . Y_k
\ea
with $i \neq j \neq k$ and 
\ba
Y_j= \left\{
\begin{array}{c} 
\bb'_j     \quad  if \quad   \delta_i =+1 \\
-\A'_j  \quad if \quad   \delta_i=-1 
\end{array} 
\right. 
\ea
It should be noted that the $Y_i$ are constructed at the point of each junction, $J$, where 
$\s_i =s_i^J(\tau)$.

Finally, energy conservation at each junction $J =\{A, B, C,. D\}$ requires that
\ba
\label{energy}
\delta_1^J \mu_1 \dot{s_1}^J + \delta_2^J \mu_2 \dot{s_2}^J 
+ \delta_3^J \mu_3 \dot{s_3}^J =0 \, .
\ea
One can check that this also follows from Eq. (\ref{s-equation}).
\section{Zipping and Unzipping}

Here we study in detail the equations of motion for $s_i(t)$. 
Let us start with junction B. At the time of collision $s_3^B(0)=0$. For the junction to form, 
$s_3^B(t)$ should be increasing initially.
For unzipping to happen, $s_3^B(t)$ should come to a stop (i.e. $\ds^B=0$) at some time
$t=t_u^B$  corresponding to the time of unzipping at junction B.
Then $s_3^B(t)$  decreases. Similarly, $s_3^D(0)=0$ initially and
after collision $s_3^D(t)$ decreases, reaching a  minimum negative value before turning back. 
The unzipping at junction D happens at $t=t_u^D $ (when $\ds^D=0$).
Interestingly, we find that $t_u^D \neq t_u^B$. The loops disentangle and separate
from each other at the time $t= t_f$ when the junctions B and D meet, corresponding to 
$s_3^D(t_f)=s_3^B(t_f)$. As we shall see, the loop disentanglement does not happen when $s_3^D(t)=s_3^B(t)=0$. It turns out that $s_3^D(t_f)=s_3^B(t_f)<0$.
Due to our symmetric construction, the same arguments go through for junctions A and C
and we can restrict the analysis to the pair of junctions B and D.

Going to the center of  mass frame, it follows from Eq. (\ref{Radius}) that
\ba
\x_{1,2}= \left(
\begin{array}{c} 
\mp b + R_0 \cos \left(\frac{t-t_0}{\gamma R_0} \right)  \cos \left(\frac{\s_{1,2}}{\gamma R_0} \right)\\
R_0 \cos \left(\frac{t-t_0}{\gamma R_0} \right)  \sin \left(\frac{\s_{1,2}}{\gamma R_0} \right)\\
\pm v t
\end{array} 
\right ) \, ,
\ea
where the impact parameter, $2b$, is the separation between the centers of the loops 
( see {\bf Fig. \ref{collision} }).
Decomposing $\x_{i}$ into left-movers as in Eq. (\ref{right-left}), yields
\ba
\A'_{1,2} = \left(
\begin{array}{c} 
 - \gamma^{-1} \sin \left(\frac{\s_{1,2} + t-t_0}{\gamma R_0} \right)  \\
\gamma^{-1} \cos \left(\frac{\s_{1,2} + t-t_0}{\gamma R_0} \right)  \\
\pm v 
\end{array} 
\right )
\quad , \quad 
\bb'_{1,2} = \left(
\begin{array}{c} 
 - \gamma^{-1} \sin \left(\frac{\s_{1,2} - t+t_0}{\gamma R_0} \right)  \\
\gamma^{-1} \cos \left(\frac{\s_{1,2} - t+t_0}{\gamma R_0} \right)  \\
\mp v 
\end{array} 
\right ) \, .
\ea

For the third string which stretches between the D and B junctions one has (following 
the arrows in {\bf Fig. \ref{collision}} where $\s_3$ increases from south to north) 
\ba
\x_{3} = ( 0, \sigma_{3}, 0)  \quad ,\quad
\A'_{3} = \bb'_{3} =  (0,1,0) \, .
\ea

 Let us start again with the junction B. Based on symmetry considerations (both loops have 
 equal tensions and radii) one expects that ${\dot s}_1^B(t)=-{\dot s}_2^B(t)$. From 
 Eq. (\ref{s-equation}) one can check that 
 ${\dot s}_1^B(t)=-{\dot s}_2^B(t)$ is a consistent solution.
 This in turn leads to $s_1^B(t) + s_2^B(t)=s_1^B(0) + s_2^B(0) $.
However, at the time of collision, $s_1^B(0)= \gamma R_0 \alpha   $ and
$s_2^B(0)= \gamma R_0 (\pi - \alpha)$ so 
\ba
\label{s12}
s_{1}^{B}(t)= - s_{2}^{B}(t) + \gamma R_{0} \pi \, .
\ea

Using the energy conservation Eq. (\ref{energy}) one obtains
\ba
\label{s23}
s_3^B(t)  = - \frac{ 2 \mu_1}{\mu_3}  \left[ s_2^B(t) - \gamma R_0 (\pi - \alpha)  \right] \,  .
\ea

The dynamical process of zipping, unzipping and loop disentanglement is 
controlled by the functions $s_3^B(t)$ and $s_3^D(t)$.
To obtain the differential equation for $\ds^B $, we first need to calculate the quantities $c_i(t)$
at junction B.  One has
\ba
c_{1} = \bb'_{2}.\bb'_{3} = \gamma^{-1}   \cos \left( \frac{s_{2}^{B}(t) -t +t_0}{\gamma R_{0}} \right)
\ea
and 
\ba
c_{2} = -\A'_{1}.\bb'_{3} = 
\gamma^{-1}   \cos \left( \frac{s_{1}^{B}(t) +t-t_0 }{\gamma R_{0}} \right) = c_{1} \, ,
\ea
where to obtain the final relation, use was made of Eq. (\ref{s12}). 
Similarly, one obtains
\ba
c_3^B = - \A_1' .  \bb_2'  = 
- 1 + 2 \gamma^{-2} \cos^{2} \left( \frac{s_{2}^{B}(t) -t+t_0 }{\gamma R_{0}} \right) \, .
\ea

With these values of $c_{i}(t)$ and using Eq. (\ref{s-equation}), one obtains
\ba
\label{junctionB}
\ds^B 
 =
\frac{   2 \mu_{1} \gamma^{-1}   \cos \left( \frac{  \mu_{3} s_{3}^{B}(t) }{2 \mu_{1} \gamma R_{0}}   + \alpha  + \frac{t-t_0}{\gamma R_0}   \right)  - \mu_{3}   }
{ 2 \mu_{1} - \mu_{3} \gamma^{-1}  \cos \left( \frac{  \mu_{3} s_{3}^{B}(t) }{2 \mu_{1} \gamma R_{0}}   + \alpha  + \frac{t-t_0}{\gamma R_0}   \right)  }  \, ,
\ea
where to get the final answer, the relation (\ref{s23}) has been used to eliminate $s_2^B(t)$ in favor
of $s_3^B(t)$.

To check the validity of the above expression, one can show that in the limit where 
$R_{0}\rightarrow \infty$, it reduces to the result of \cite{Copeland:2006eh}
for collision of two infinite straight strings.

Following the same steps, for the junction D one finds
\ba
\label{junctionD}
\ds^D= -\frac{   2 \mu_{1} \gamma^{-1}   \cos \left( \frac{  \mu_{3} s_{3}^{D}(t) }{2 \mu_{1} \gamma R_{0}}   + \alpha  - \frac{t-t_0}{\gamma R_0}   \right)  - \mu_{3}   }
{ 2 \mu_{1} - \mu_{3} \gamma^{-1}  \cos \left( \frac{  \mu_{3} s_{3}^{D}(t) }{2 \mu_{1} \gamma R_{0}}   + \alpha  - \frac{t-t_0}{\gamma R_0}   \right)  }  \, .
\ea
Comparing the equations for $\ds^B$ and $\ds^D$, we observe that 
$s_3^B \rightarrow -s_3^D $ under time reversal $t-t_0 \rightarrow -(t- t_0)$.

One can check that for the junctions A and B the evolution of $\ds^A$ and $\ds^C$ are  
identical to that of $\ds^B$ and $\ds^D$, with a sign difference as expected due to our symmetric construction.

With some effort, one can solve Eqs. (\ref{junctionB}) and (\ref{junctionD}) with the answer
\ba
\label{s3B}
\frac{s_3^B}{R_0}  - \sin \left( \frac{  \mu_{3} s_{3}^{B} }{2 \mu_{1} \gamma R_{0}}   + \alpha  + \frac{t-t_0}{\gamma R_0}   \right) =   
- \sin \left(\alpha -  \frac{t_0}{ \gamma R_0} \right)  - \frac{ \mu_3}{2 \mu_1 R_0} t  \, ,
\ea
which expresses $s_3^B(t)$ implicitly as a function of $t$. A similar equation holds
for $s_3^D$ with $(t, t_0)  \rightarrow -( t, t_0)$.

The above implicit equations for  $s_3^B$ and $s_3^D$ can not be solved explicitly to obtain
the variables as functions of $t$.
However, some insight can be obtained by looking at the form of Eq. (\ref{junctionB}). 
For the junction B to materialize at $t=0$, we need that $\dot s_3^B(0) >0$. 
This requires that $\gamma \mu_3 < 2 \mu_1 $ and
\ba
\label{angle1}
\left( \alpha -  \frac{t_0}{ \gamma R_0} \right) < \alpha_c \equiv  \cos^{-1} \left( \frac{\mu_3 \gamma}{2 \mu_1}   \right) \, .
\ea
Interestingly, this is the same junction formation condition  as for the collision of straight 
strings \cite{Copeland:2006eh}  where $R_{0}\rightarrow \infty$. This is understandable, since 
the collision and junction formation is a local
effect and at the points of collision large loops may be approximated as straight strings.
On the other hand, for the junction D to materialize after collision, one expects that 
$\dot s_3^D(0) <0$ which yields
\ba
\label{angle2}
\left( \alpha +  \frac{t_0}{ \gamma R_0} \right) < \alpha_c
\ea
Interestingly, when $t_0 \neq 0$, Eq. (\ref{angle2}) is stronger a condition than Eq. (\ref{angle1}). 

\begin{figure}[t] 
   \centering
   \includegraphics[width=2.2in]{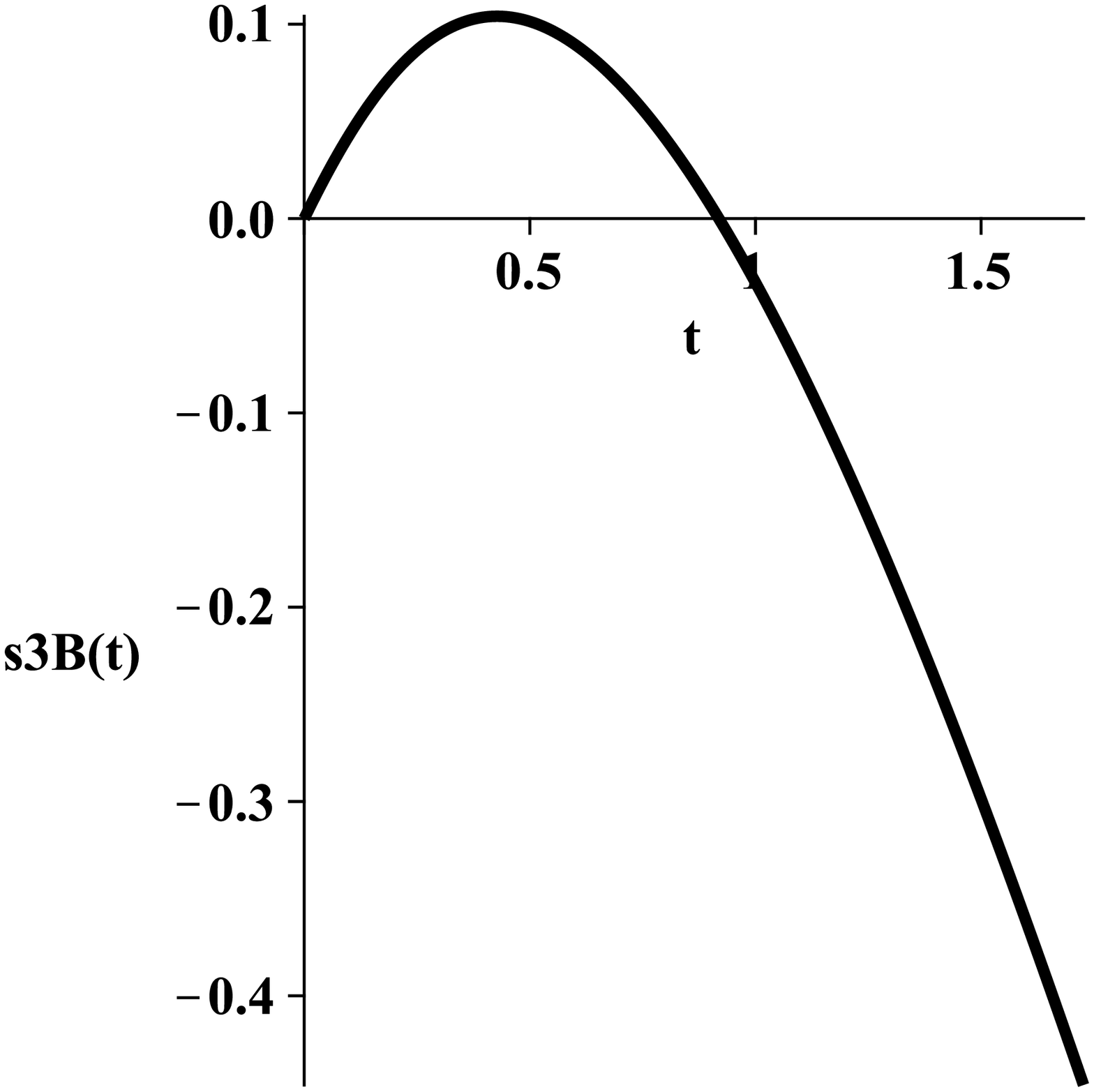} \hspace{2cm}
     \includegraphics[width=2.2in]{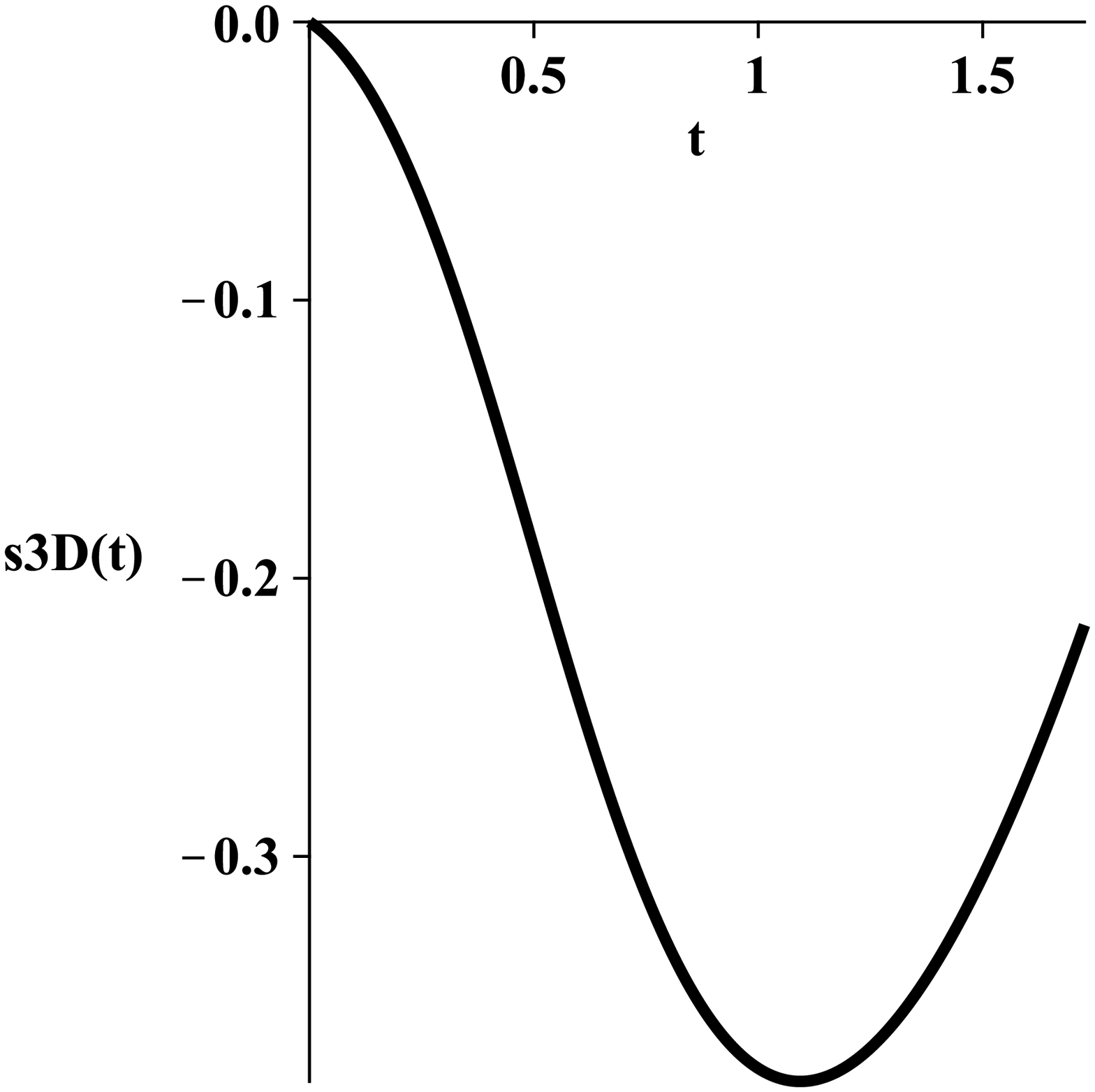} 
   \hspace{0cm}
\caption{ Here the numerical solutions for $s_3^B(t)$(left) and $s_3^D$(right) are plotted for $\gamma=1.1, \mu_1/\mu_3=0.7, \alpha= \pi/8$
and $t_0=0.1$ in units where $R_0=1$. The  unzipping for junction B(D) happens when
$s_3^B$($s_3^D$) reaches a maximum(minimum) value. In this example, the loop 
disentanglement happens at $t_f \simeq1.4$ before the loops shrink at 
$t_{shrink} \simeq 1.7$.
}

\label{unzip}
\end{figure}

Once the junction B is formed, it grows until the time  $t_u^B$ of unzipping, when the 
argument inside the $\cos$ function in Eq. (\ref{junctionB}) becomes equal
to $\alpha_c$ and $\ds^B=0$. As time goes by,  the argument inside the $\cos$ function 
increases, $\ds^B$ becomes negative and the junction B turns back. A similar argument 
applies to junction D except that the unzipping happens at the time $t=t_u^D$, and due to 
the time asymmetry in Eqs. (\ref{junctionB}) and (\ref{junctionD}), $t_u^D \neq t_u^B$. Below we 
will demonstrate that $t_u^D > t_u^B$.
After  $ t > t_u^D$, the junctions B and D move towards each other. The loops disentangle 
at the time  $t=t_{f}$ when the junctions meet, corresponding to $s_3^B(t_f) = s_3^D(t_f)$. 
In {\bf Fig. \ref{unzip}} we have plotted the shapes of $s_3^B$ and $s_3^D$  for some given 
parameter values of $\alpha, \gamma, \mu_1, \mu_2$ and $R_0$.
The left figure indicates that $s_3^B$($s_3^D$) increases(decreases) initially and then come to a 
halt,  indicating the time of unzipping. 

Here we would like to find the time of unzipping and loop disentanglement. Consider junction B. 
At the time $t = t_u^B$ of unzipping one obtains from $\ds^B=0$ that
\ba
\frac{s_3^B(t_u^B)}{R_0} = \frac{2 \mu_1 \gamma}{ \mu_3}
\left(  \alpha_c -  \alpha  -  \frac{ t_u^B- t_0}{ \gamma R_0}\right) \, .
\ea
Plugging this into Eq. (\ref{s3B}) gives the unzipping time
\ba
\label{tuB}
\frac{t_u^B}{R_0} = 
\left( 1- \frac{\mu_3^2}{4 \mu_1^2}    \right)^{-1}
\left\{
   \gamma \left(  \alpha_c -  \alpha \right) +  \frac{  t_0}{ R_0}
+ \frac{\mu_3}{2 \mu_1} \left[ \,   \sin (\alpha -  \frac{t_0}{ \gamma R_0} )
- \sqrt{1- \frac{ \gamma^2 \mu_3^2 }{4 \mu_1^2} \,   } \, 
\right]   \right\} \, .
\ea

To find the unzipping time for junction D, we note that after junction formation the argument inside the cos function in Eq. (\ref{junctionD}) decreases with time. It becomes negative and the
unzipping for junction D happens when the expression inside the cos function becomes equal to 
$- \alpha_c$. With this consideration and following the steps as above yields 
\ba
\label{tuD}
\frac{t_u^D}{R_0} = 
\left( 1- \frac{\mu_3^2}{4 \mu_1^2}    \right)^{-1}
\left\{
   \gamma \left(   \alpha_c +  \alpha \right) +  \frac{  t_0}{ R_0}
- \frac{\mu_3}{2 \mu_1} \left[ \,   \sin (\alpha +  \frac{t_0}{ \gamma R_0} )
+ \sqrt{1- \frac{ \gamma^2 \mu_3^2 }{4 \mu_1^2} \,   } \, 
\right]   \right\} \, .
\ea

Equations (\ref{tuB}) and (\ref{tuD}) are implicit equations which relate $t_u^B$ and $t_u^D$
to the tensions $\mu_i$, the incoming angle of collision $\alpha$, the velocity $\gamma$ and the 
initial loop phase $t_0$. It is not easy to see how $t_u^B$ and $t_u^D$ vary as one varies 
these parameters simultaneously. As a simple treatment, let us take $\mu_i$ and $\gamma$ as 
fixed properties of a network of cosmic strings and consider the unzipping times as functions of 
$\alpha$ and $t_0$ (which may be considered as random parameters for the network evolution).
If one increases $t_0>0$ while keeping $\alpha $ fixed, then both $t_u^B$ and $t_u^D$ increase.
There is a limit on how large $t_0$ can be. This is determined by Eq. (\ref{angle2}).
The dependence of the unzipping on $\alpha$ is more non-trivial. From Eqs.  (\ref{tuB}) and 
(\ref{tuD}) we note that the dependence of these times on $\alpha$ is not symmetric. For a fixed 
value of $t_0$, then as $\alpha$ increases, $t_u^D$ increases while $t_u^B$ decreases almost 
linearly with $\alpha$. Again, there is a limit on how big $\alpha$ can be, which is determined by 
Eq. (\ref{angle2}).

It is instructive to see which of the junctions B or D starts to unzip sooner. From the above two 
equations, the difference in the unzipping times is calculated to be
\ba
\frac{t_u^D- t_u^B}{R_0} = 2 \gamma  \alpha \left( 1- \frac{\mu_3^2}{4 \mu_1^2}    \right)^{-1}
\left[ 1 - \frac{\mu_3}{2 \mu_1 \gamma} \cos(\frac{t_0}{\gamma R_0})  \, 
\frac{\sin \alpha}{\alpha} \,
\right] \, .
\ea
Since $\sin \alpha/\alpha$
is always less than unity we see that $t_u^D > t_u^B$. This means that the junction B which holds 
the external large arcs unzips sooner than the junction D which holds the internal small arcs. 
Keeping all other parameters fixed, by increasing the angle of collision $\alpha$, the difference 
in unzipping times increases almost linearly with $\alpha$.

The time $t_{f}$ of loop disentanglement is given by $s_3^B(t_{f})=s_3^D(t_{f})$. Using 
Eq. (\ref{s3B}) and the similar equation for $s_3^D$ gives
\ba
\frac{2 \gamma \mu_1 }{\mu_3} \cos^{-1} \Gamma -  \cos\left(\frac{t-t_0}{\gamma R_0}\right)
\sqrt{1-\Gamma^2} - \frac{2 \gamma \mu_1 \alpha }{\mu_3}  + \sin \alpha \, \cos(\frac{t_0}{\gamma R_0}) =0 \, ,
\ea
where
\ba
\Gamma \equiv \left[  \frac{ \mu_3 t /(2 \mu_1 R_0)  - \cos \alpha \sin(\frac{t_0}{\gamma R_0}) }{ \sin(\frac{t-t_0}{\gamma R_0})}
 \right] \, .
\ea
This is an implicit equation for $t_{f}$ which should be solved in terms of $\mu_i, \gamma, \alpha, t_0$ and $R_0$. For this to make sense, we demand that 
$t_{f} -t_0 <  \pi\, R_0/2 $ before the loops shrink to zero.

\section{Discussion}

In this paper  we have provided a theoretical understanding of the zipping-unzipping 
phenomena in cosmic string loop collisions.
The process of unzipping and string disentaglement has important effects on the evolution of 
networks containing different types of strings. Initially, one may fear that  the over-abundance of
junctions and the string bound states may lead to a frustrated network of cosmic strings, 
preventing the network to reach a scaling regime. In an interesting simulation run by Urrestilla 
and Vilenkin \cite{Urrestilla:2007yw}
it was shown that the presence of junctions and bound states is not catastrophic. Indeed, it was 
shown that for a network containing two different types of strings, the contribution of the bound 
states  to the population and length is negligible compared to that 
of the original strings. There may be two reasons for why the contribution of the junctions and bound states to the network's string length and  number density is sub-dominant. Firstly,  
cosmic strings move with very high
velocities and can simply pass through each other, and no junctions form in the first 
place \cite{Copeland:2006eh, Copeland:2006if, Copeland:2007nv, Achucarro:2006es}. Secondly, and more 
curiously, junctions may materialize occasionally but  they soon become unstable to unzipping. 
This was the subject of our current study.
 
To simplify the analysis, here  we considered the simple case when the colliding loops have 
equal tensions and radii. In principle one can consider more general cases when loops have 
different tensions and configurations. 

In examples of straight strings in collision 
\cite{Copeland:2006eh, Copeland:2006if, Copeland:2007nv}, the junctions do not stop growing in 
time once they are formed.
In contrast, we have demonstrated that for colliding loops unzipping phenomena take place . 
It is energetically costly for junctions to grow indefinitely. 
The junctions holding the external loops and those holding the internal loops behave differently. 
The junctions holding the external large loops start to unzip sooner than the junctions holding 
the internal small ones.  The onset of unzipping and eventual loops disentanglement is 
determined by the parameters of the colliding loops such as their tensions, the angle of 
collision and their velocity.

\section{Acknowledgments}

We would like to thank  T. Kibble, R. Myers, S. Sheikh-Jabbari, T. Vachaspati  and
A. Vilenkin for useful discussions and comments.
H.F., J.K. and R.B. would like to acknowledge the hospitality of the KITPC, where this work was started, 
and the Perimeter Institute, where this work was finished.
At McGill, this research has been supported by NSERC
under the Discovery Grant program. R.B. is also
supported by the Canada Research Chairs program.



\end{document}